\documentstyle[amssymb,psfig]{mn}

\title[Star Clusters in Starburst Galaxies. I.]{Star Cluster Formation and
Evolution in Nearby Starburst Galaxies: I.  Systematic Uncertainties}

\author[R.  de Grijs et al.]{R. de Grijs$^1$\thanks{E-mail:
grijs@ast.cam.ac.uk}, U. Fritze--v. Alvensleben$^2$, P. Anders$^2$,
J.S. Gallagher {\sc iii}$^3$, 
\newauthor N. Bastian$^4$, V.A. Taylor$^5$, R.A. Windhorst$^5$ \\
$^1$ Institute of Astronomy, University of Cambridge, Madingley Road,
Cambridge CB3 0HA \\
$^2$ Universit\"atssternwarte, University of G\"ottingen,
Geismarlandstr. 11, 37083 G\"ottingen, Germany \\
$^3$ Astronomy Department, University of Wisconsin-Madison, 475 N.
Charter St., Madison, WI 53706, USA \\
$^4$ Astronomical Institute, Utrecht University, Princetonplein 5,
3584 CC Utrecht, The Netherlands \\
$^5$ Department of Physics \& Astronomy, Arizona State University, Box
871504, Tempe, AZ 85287-1504, USA
}

\date{Accepted ---. Received ---; in original form ---.}

\pubyear{2002}

\begin{document}
\maketitle

\def\asec {{$\buildrel{\prime\prime}\over .$}}

\begin{abstract}
The large majority of extragalactic star cluster studies done to date
have essentially used two or three-passband aperture photometry,
combined with theoretical stellar population synthesis models, to obtain
age, mass and extinction estimates, and sometimes also metallicities. 
The accuracy to which this can be done depends on the choice of
(broad-band) passband combination and, crucially, also on the actual
wavelengths and the wavelength range covered by the observations. 
Understanding the inherent systematic uncertainties (the main aim of
this paper) is of the utmost importance for a well-balanced
interpretation of the properties of extragalactic star cluster systems. 
\\
We simultaneously obtain ages, metallicities and extinction values for
$\sim 300$ clusters in the nearby starburst galaxy NGC 3310, based on
archival {\sl Hubble Space Telescope} observations from the ultraviolet
(UV) to the near-infrared (NIR).  We show that, for ages $6 \lesssim
\log( {\rm age/yr} ) \lesssim 9$, and if one can only obtain partial
coverage of the spectral energy distribution (SED), an optical passband
combination of at least four filters including {\it both} blue {\it and}
red passbands results in the most representative age distribution, as
compared to the better constrained ages obtained from the full UV--NIR
SED coverage.  We find that while blue-selected passband combinations
lead to age distributions that are slightly biased towards younger ages
due to the well-known age--metallicity degeneracy, red-dominated
passband combinations should be avoided.  \\
NGC 3310 underwent a (possibly extended) global burst of cluster
formation $\sim 3 \times 10^7$ yr ago.  This coincides closely with the
last tidal interaction or merger with a low-metallicity galaxy that
likely induced the formation of the large fraction of clusters with
(significantly) subsolar metallicities.  The logarithmic slope of the
{\it V}-band cluster luminosity function, for clusters in the range
$17.7 \lesssim {\rm F606W} \lesssim 20.2$ mag, is $\alpha_{\rm F606W}
\simeq -1.8 \pm 0.4$.  The observed cluster system has a median mass of
$\langle \log( m/M_\odot ) \rangle \simeq 5.25 \pm 0.1$, obtained from
scaling the appropriate model SEDs for known masses to the observed
cluster SEDs. 
\end{abstract}


\section{The Cluster Luminosity Function}

\subsection{Introduction}
\label{intro.sec}

Cluster luminosity functions (CLFs) and colour distributions are the
most important diagnostics in the study of globular and compact star
cluster populations in nearby galaxies.  For the old globular cluster
(GC) systems in, e.g., the Galaxy, M31, M87, and old elliptical
galaxies, the CLF shape is well-established: it is roughly Gaussian,
with the peak or turnover magnitude at $M_V^0 \simeq -7.4$ mag and a
Gaussian FWHM of $\sim 3$ mag (Whitmore et al.  1995, Harris 1996, 2001,
Ashman \& Zepf 1998, Harris, Harris \& McLaughlin 1998).  The
well-studied young star cluster (YSC) population in the LMC and the
Galactic open cluster system, on the other hand, display a power-law CLF
(Elson \& Fall 1985, Harris \& Pudritz 1994, Elmegreen \& Efremov 1997). 

{\sl Hubble Space Telescope (HST)} observations are continuing to
provide an ever increasing number of CLFs for compact YSC systems in
galaxies beyond the Local Group and the Virgo cluster (e.g., Whitmore \&
Schweizer 1995, Schweizer et al.  1996, Miller et al.  1997, Zepf et al. 
1999, de Grijs, O'Connell \& Gallagher 2001, Whitmore et al.  2002). 
Although a large number of studies have attempted to detect a turn-over
in young or intermediate-age CLFs, the shapes of such young and
intermediate-age CLFs have thus far been consistent with power laws down
to the observational completeness thresholds (but see Miller et al. 
1997, de Grijs et al.  2001, 2003a,b). 

\subsection{Evolutionary Effects?}

The striking difference between the power-law distributions for young
star clusters and the Gaussian distribution of the old Galactic GCs has
recently attracted renewed theoretical attention.  The currently most
popular (but, admittedly, speculative) GC formation models suggest that
the distribution of the initial cluster masses is closely approximated
by a power law (e.g., Harris \& Pudritz 1994, McLaughlin \& Pudritz
1996, Elmegreen \& Efremov 1997, Gnedin \& Ostriker 1997, Fall \& Zhang
2001). 

Which processes will affect the CLFs such that they transform from a
power-law shape to a Gaussian distribution? It is generally assumed that
the processes responsible for the depletion of a star cluster population
over time-scales of a Hubble time include the preferential depletion of
low-mass clusters both by evaporation due to two-body relaxation and by
tidal interactions with the gravitational field of their host galaxy
(e.g., Fall \& Rees 1977, 1985, Elmegreen \& Efremov 1997, Murali \&
Weinberg 1997a,b,c, Ostriker \& Gnedin 1997, Harris et al.  1998, Fall
\& Zhang 2001), and the preferential disruption of high-mass clusters by
dynamical friction (Vesperini 2000, 2001).  From the currently most
popular GC evolution models it follows that {\it any} initial mass (or
luminosity) distribution will shortly be transformed into a peaked
distribution, although it should be noted that these models apply only
to Milky Way-type conditions in which the GC system is characterised by
significant radially dependent radial anisotropy.  Vesperini (2000,
2001) has included the internal gravitational interactions between
cluster stars in his models and concludes that these need considerable
fine tuning to transform a power law initial cluster mass function
(ICMF) into a Gaussian distribution, whereas a Gaussian ICMF conserves
its shape rather independently of the choice of parameters: destruction
of low-mass clusters by evaporation and the tidal field is balanced by
the destruction of high-mass clusters through dynamical friction. 

\subsection{Interacting galaxies}

All of these models are valid {\it only} for Milky Way-type
gravitational potentials, however.  Galaxy-galaxy interactions will
obviously have a major effect on the resulting gravitational potential,
in which the dynamical star cluster evolution is likely significantly
different (see, e.g., Boutloukos \& Lamers 2003, de Grijs et al. 
2003a). 

The mass spectrum of molecular clouds or molecular cloud cores,
progenitors of young star clusters, has never been determined in
interacting galaxies to similarly faint limits as in the Milky Way and
the Magellanic Clouds.  The significant external pressure in interacting
gas-rich galaxies may be expected to have a major effect on the
molecular cloud mass spectrum. 

Significant age spread effects in young cluster systems -- in which
cluster formation is still ongoing -- affect the observed CLF (Meurer
1995, Fritze--v.  Alvensleben 1998, 1999, de Grijs et al.  2001,
2003a,b), which might in fact make an intrinsically Gaussian CLF (or
ICMF) appear as a power-law CLF (see, e.g., Miller et al.  1997,
Fritze--v.  Alvensleben 1998).  It is obviously very important to age
date the individual clusters and to correct the observed CLF to a common
age, before interpreting the CLF (Fritze--v.  Alvensleben 1999, de Grijs
et al.  2001, 2003a,b), in particular if the age spread within a cluster
system is a significant fraction of the system's mean age. 

In de Grijs et al.  (2003a,b), we provide the first observational
evidence for a clear turn-over in the intermediate-age CLF of the
clusters formed {\it in the burst of cluster formation} in M82's fossil
starburst region ``B'', which very closely matches the universal CLFs of
old GC systems.  This provides an important test of cluster disruption
models (see de Grijs et al.  2003b). 

\subsection{Star cluster metallicities}

Metallicities of YSCs produced in galaxy interactions, mergers or
starbursts are an important discriminator against GCs formed in the
early Universe.  They are expected to correspond to the interstellar
medium (ISM) abundances of the interacting/starburst galaxies, and are
therefore most likely significantly higher than those of halo GCs in the
Milky Way and other galaxies with old GC systems.  ISM abundances span a
considerable range, however, among different galaxy types from Sa
through Sd, irregular, and dwarf galaxies, and may exhibit significant
radial gradients (Oey \& Kennicutt 1993, Zaritsky, Kennicutt \& Huchra
1994, Richer \& McCall 1995).  Many of these galaxies show abundance
gradients, which have recently been shown to sometimes extend
considerably beyond their optical radii (Ferguson et al.  1998, van Zee
et al.  1998).  However, from those large radii, gas can be funneled
efficiently into the inner regions during galaxy interactions and
mergers (e.g., Hibbard \& Mihos 1995). 

Hence, a considerable metallicity range may be expected for YSCs
produced in interactions of various types of galaxies and even among the
YSCs formed within one global galaxy-wide starburst. 

During a strong burst, typically lasting a few $\times 10^8$ yr in a
massive gas-rich galaxy, a significant increase of the ISM abundance may
occur (Fritze--v.  Alvensleben \& Gerhardt 1994, their Fig.  12b). 
Meanwhile, some fraction of the gas enriched by dying first-generation
burst stars may well be shock-compressed to cool fast enough to be built
into later generations of stars or clusters produced in the burst.  The
same effect may occur when multiple bursts occur in a series of close
encounters between two galaxies before their final merger.  Hence, in
extended starburst episodes metallicity differences between YSCs formed
early on and late in the burst phase may be expected. 

Precise (relative) metallicity determinations for individual YSCs are
not only important to study the questions raised above, but also for the
correct derivation of ages from broad-band colours. 

Dust extinction is often very important in YSC systems.  In particular
the youngest post-burst galaxies and galaxies with ongoing starbursts
often show strong and patchy dust structures and morphologies.  For
instance, the youngest clusters in the overlap region of the two
galactic discs in the Antennae galaxies are completely obscured in the
optical and can only be detected in near or mid-infrared observations
(Mirabel et al.  1998, Mengel et al.  2001).  Older merger remnants like
NGC 7252 or NGC 3921 seem to have blown their inner regions clear of all
the gas left over from intense star formation during the burst (e.g.,
Schweizer et al.  1996).  Extinction estimates towards individual YSCs
are therefore as important as individual metallicity estimates in order
to obtain reliable ages and to be able to derive an age-normalised CLF
or YSC mass function. 

\subsection{Spectroscopy vs. multi-passband photometry}

Individual YSC spectroscopy, feasible today with 8m-class telescopes for
the nearest systems, is very time-consuming, since observations of large
numbers of clusters are required to obtain statistically significant
results.  Multi-passband imaging is a very interesting and useful
alternative, as we will show below, in particular if it includes
coverage of near-infrared (NIR) and/or ultraviolet (UV) wavelengths. 
The large majority of extragalactic star cluster studies done to date
have essentially used two or three-passband aperture photometry,
combined with theoretical stellar population synthesis models to obtain
age estimates.  The accuracy to which this can be done obviously depends
on the number of different (broad-band) filters available as well as,
crucially, on the actual wavelengths and wavelength range covered by the
observations, and on the PSF size compared to the cluster surface
density profile (i.e., on how close to the observations were taken to
the confusion limit for these clusters). 

In this paper we assess the systematic uncertainties in age, extinction
and metallicity determinations for YSC systems inherent to the use of
broad-band, integrated colours.  We have developed an evolutionary
synthesis optimisation technique that can be applied to photometric
measurements in a given number $N (N \ge 4)$ of broad-band passbands. 
The optimisation routine then simultaneously determines the best
combination of age, extinction and metallicity from a comparison with
the most up-to-date G\"ottingen simple stellar population (SSP) models
(Schulz et al.  2002), to which we have added the contributions of an
exhaustive set of gaseous emission lines and gaseous continuum emission
(Anders, Fritze--v.  Alvensleben \& de Grijs 2002, Anders \& Fritze--v. 
Alvensleben 2003).  We also compare these results with similar
determinations based on the Starburst99 SSP models (Leitherer et al. 
1999), but assuming fixed, solar metallicity for our sample clusters. 
Although this is an often-used assumption, we will show that this
introduces significant systematic effects in the final age distribution,
and therefore also in the mass distribution. 

We decided to focus our efforts on the nearby, well-studied starburst
galaxy NGC 3310, known to harbour large numbers of young star clusters,
for which multi-passband observations from the near-UV to the NIR are
readily available from the {\sl HST} Data Archive (Section
\ref{obs.sect}; see also Elmegreen et al.  2002, hereafter E02).  In
Section \ref{ngc3310.sect} we first place the NGC 3310 starburst in its
physical context.  We then discuss the derived age distribution of the
cluster population, which we extend compared to previous work, in terms
of the evolution of the CLF and the interaction stage of its parent
galaxy in Sections \ref{agedet.sec} and \ref{interpretation.sec}.  We
summarise our main results and conclusions in Section
\ref{summary.sect}.  Finally, we will apply our knowledge of the
systematic uncertainties gained in this paper to a larger sample of
nearby starburst and interacting galaxies drawn from the {\sl HST}
GO-8645 programme (Windhorst et al.  2002) in Papers II and III (de
Grijs et al., in prep.). 

\section{Observations and data preparation}
\label{obs.sect}

NGC 3310 is a representative member of the class of galaxies often
showing signs of active starbursts and recently formed star clusters. 
The galaxy may have been a normal, quiescent Sbc-type galaxy before it
started to produce large numbers of new stars, possibly due to the
merger with a gas-rich low-metallicity companion (cf.  Kobulnicky \&
Skillman 1995).

As part of {\sl HST} programme GO-8645, we obtained observations of NGC
3310 through the F300W (``UV'') and F814W filters (Windhorst et al. 
2002), with the galaxy centre located on chip 3 of the Wide Field
Planetary Camera 2 ({\sl WFPC2}).  Observations in additional passbands
were obtained from the {\sl HST} Data Archive.  In order to obtain the
largest common field of view (FoV) in the optical wavelength range, we
restricted these archival data to be taken with the {\sl WFPC2}.  In
addition, we obtained archival NIR images taken with the Near-Infrared
Camera and Multi-Object Spectrometer's (NICMOS) camera 2, which provides
the best compromise of spatial resolution (0\asec075 pixel$^{-1}$) and
FoV (19\asec2$\times$19\asec2) when combined with {\sl WFPC2}
observations (see below).  We have summarized the combined set of {\sl
HST} observations through broad-band filters used in this paper in Table
\ref{obs.tab}. 

\begin{table}
\caption[ ]{\label{obs.tab}Overview of the {\sl HST} observations of NGC
3310
}
{\scriptsize
\begin{center}
\begin{tabular}{crccr}
\hline
\hline
\multicolumn{1}{c}{Filter} & \multicolumn{1}{c}{Exposure time} &
\multicolumn{1}{c}{Centre$^a$} & \multicolumn{1}{c}{PID$^b$} &
\multicolumn{1}{c}{ORIENT$^c$} \\
& \multicolumn{1}{c}{(sec)} & & & \multicolumn{1}{c}{($^\circ$)} \\
\hline
F300W & 900,1000        & WF3  & 8645 & $-$150.282 \\
F336W & 2$\times$500    & PC   & 6639 & $-$144.232 \\
F439W & 2$\times$300    & PC   & 6639 & $-$144.232 \\
F606W & 500             & PC   & 5479 &    145.553 \\
F814W & 160,180         & PC   & 6639 & $-$144.232 \\
      & 100,160         & WF3  & 8645 & $-$150.282 \\
F110W & 2$\times$159.96 & NIC2 & 7268 &     80.519 \\
F160W & 2$\times$191.96 & NIC2 & 7268 &     80.519 \\
\hline
\end{tabular}
\end{center}
{\sc Notes:} $^a$ -- Location of the galactic centre; $^b$ -- {\sl HST}
programme identifier; $^c$ -- Orientation of the images (taken from the
image header), measured North through East with respect to the V3 axis
(i.e., the X=Y diagonal of the WF3 CCD $+ 180^\circ$). 
}
\end{table}

Pipeline image reduction and calibration of the {\sl WFPC2} images were
done with standard procedures provided as part of the {\sc
iraf/stsdas}\footnote{The Image Reduction and Analysis Facility (IRAF)
is distributed by the National Optical Astronomy Observatories, which is
operated by the Association of Universities for Research in Astronomy,
Inc., under cooperative agreement with the National Science Foundation. 
{\sc stsdas}, the Space Telescope Science Data Analysis System, contains
tasks complementary to the existing {\sc iraf} tasks.  We used Version
2.2 (August 2000) for the data reduction performed in this paper.}
package, using the updated and corrected on-orbit flat fields and
related reference files most appropriate for the observations. 

We first registered the individual images obtained for a given programme
and created combined images for each programme/passband combination,
using the appropriate {\sc imalign} and {\sc imcombine} routines in {\sc
iraf}.  Next, we rotated all of these combined images to a single
orientation using objects in common among the observed fields, applying
{\sc iraf}'s {\sc rotate} and {\sc imgeom} routines.  We used the
observations of programme GO-6639 as basis for these rotations. 
Finally, we adjusted the pixel size of both the PC and the NICMOS/2
images to that of the final, rotated WF3 images.  The final pixel size
for all images is 0\asec0998. 

Because of the significantly smaller FoV of the NICMOS observations, we
created two sets of final, registered images: one set containing the
{\sl WFPC2} images only, and a second set consisting of all aligned
images ({\sl WFPC2} and NICMOS).  The resulting combined ``{\sl
WFPC2}-only'' FoV is $331 \times 323$ pixels (33\asec03 $\times$
32\asec24).  Similarly, the common FoV for all instrument/passband
combinations is $226 \times 227$ pixels (22\asec55 $\times$ 22\asec65). 

\subsection{Source selection}
\label{sources.sect}

We based our initial selection of sources on a customized version of the
{\sc daofind} task in the {\sc daophot} software package (Stetson 1987),
as running under {\sc idl}.

We obtained initial source lists for all available passbands, using a
detection limit of four times the r.m.s.  noise in the (global) sky
background, determined from the individual images.  We did not force our
detection routine to constrain the source roundness or sharpness, in
order to be as inclusive as possible. 

For NGC 3310, at a distance $D \simeq 13$ Mpc (Bottinelli et al.  1984,
Kregel \& Sancisi 2001), most star clusters and star forming regions
appear as point-like sources, and the detection of individual stars
among these is unlikely given the associated high luminosities required. 
Thus, we can use the initial source lists as a starting point to obtain
the final, verified list of candidate star clusters, as outlined below. 

We cross-correlated the source lists obtained in each pair of subsequent
passbands, allowing for only a 1.5-pixel positional mismatch between the
individual images.  The F606W $\otimes$ F814W cross-correlated source
lists were subsequently adopted as the basis for our final source lists,
to which we added the source detections resulting from the cross
correlations of the other passband pairs, if not already included.

To reject artefacts remaining in the final source lists and real sources
that were badly situated for aperture photometry, we visually examined
all of the candidates on enlargements of both of the images from which
they originated for contrast, definition, aperture centering, and
background sampling.  We rejected candidates that were too diffuse or
that might be the effects of background fluctuations.  A small number of
sources contained multiple components or substructure (as expected for
young stellar associations), and the apertures were adjusted to include
all of these.  We also visually inspected the 2-D Gaussian fits to those
sources that were assigned $\sigma_{\rm Gauss} \ge 2$ pixels in the
initial automated fitting pass.  If needed, apertures for our final
photometric pass (and thus the source magnitudes), centre coordinates,
and $\sigma_{\rm Gauss}$'s were adjusted.  The visual verification is a
justifiable step in the reduction process, since -- after all -- we are
mainly interested in obtaining good photometry for a representative,
sample of (young) star clusters, for which we can reliably derive the
systematic uncertainties in their age, extinction and metallicity
determinations. Hence, we sacrifice some sample completeness in order to
achieve higher sample reliability.

Thus, our primary cluster candidate samples consist of well-defined
sources with $\sigma_{\rm Gauss} \ge 1.20$ pixels and relatively smooth
backgrounds.  The total number of visually verified sources contained in
our source lists is 382 and 289, respectively, for the ``{\sl
WFPC2}-only'' FoV and for the combined FoV. 

\subsection{Photometric calibration}

The coordinates from the source lists obtained in the previous section  
were used as the centres for {\sc daophot} aperture photometry in all  
passbands.

The correct choice of source and background aperture sizes is critical
for the quality of the resulting photometry.  Due to the complex
structure of the star-forming regions, we concluded that we had to
assign apertures for source flux and background level determination
individually to each cluster candidate by visual inspection.  The
``standard'' apertures for the majority of the sources were set at a
5-pixel radius for the source aperture and an annulus between 5 and 8
pixels for the background determination, although in individual cases we
had to deviate significantly from these values (e.g., because of
extended source sizes, contamination by nearby objects, or gradients in
the local sky background).  Our photometry includes most, if not all, of
the light of each cluster candidate. 

The photometric calibration, i.e., the conversion of the instrumental
aperture magnitudes thus obtained to the {\sl HST}-flight system
(STMAG), was done by simply using the appropriate zero-point offsets for
each of the individual passbands, after correcting the instrumental
magnitudes for geometric distortions, charge transfer (in)efficiency
effects, and the exposure times, following procedures identical to those
applied in de Grijs et al.  (2002). 

The full data tables containing the integrated photometry of all
clusters are available on-line, at {\tt
http://www.ast.cam.ac.uk/STELLARPOPS/Starbursts/}. 

Special care needs to be taken when calibrating the UV aperture
magnitudes, in particular the F300W and F336W fluxes.  These filters
suffer from significant ``red leaks'' (Biretta et al.  2000, chapter 3). 
This causes a fraction of an object's flux longward of $\sim 6000${\AA}
to be detected in these filters.  Close inspection of the filter
transmission curves reveals that the response curve of the red-leak
region of the F300W filter resembles the transmission curve of the F814W
filter, being most dominant in the 7000--9000{\AA} range.  The red leak
of the F336W filter is roughly similar to the red half of the F814W
filter response function.  This is fortunate, and implies that we could
in principle use our F814W observations to correct the F300W and F336W
fluxes for red-leak contamination.  Biretta et al.  (2000) show that the
red leak in these two filters is in general $\lesssim 5$\% of the total
flux for stellar populations dominated by K3V or earlier-type stars,
although it can be as much as 10--15\% of the total flux for stellar
populations dominated by M0V--M8V stars.  Thus, for starburst galaxies
dominated by young hot stellar populations the red leak contamination
should be almost negligible, typically $\lesssim 1$\%. 

For realistic spectral energy distributions (SEDs) for early-type
galaxies (dominated by older stellar populations), we have shown that
the F300W red leak is typically 5--7\% of the total F300W flux, and
never exceeds the 10\% level, not even in the reddest galactic bulges
(Windhorst et al.  2002, Eskridge et al.  2003).  Adopting Eskridge et
al.'s (2003) most straightforward assumption, namely that the red leak
cannot account for any more than all of the observed counts in any given
area of a few pixels in the F300W and/or F336W images, we derive a
maximum contribution of the red leak in NGC 3310 to both the F300W and
the F336W filters of 3\% of the total count rate {\em in the F814W
image}.  We applied these corrections to the F300W and F336W images
before obtaining the aperture photometry. 

\subsection{Completeness analysis}

\begin{figure}
\psfig{figure=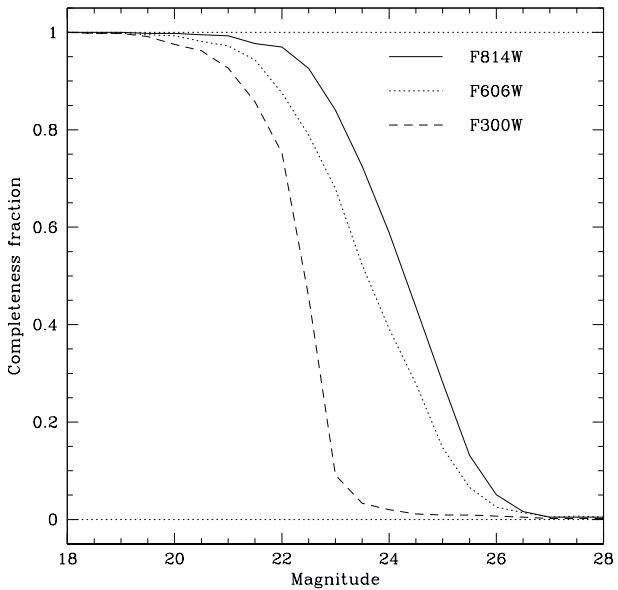,width=8.5cm}
\caption{\label{compl.fig}Completeness curves for the {\sl WFPC2} FoV of
NGC 3310.  The different line styles refer to different passbands, as
indicated in the figure.}
\end{figure}

We estimated the completeness of our source lists by using synthetic
source fields consisting of PSFs.  We created artificial source fields
for input magnitudes between 18.0 and 28.0 mag, in steps of 0.5 mag,
independently for each of the F300W, F606W, and F814W passbands.  We
then applied the same source detection routines used for our science
images to the fields containing the combined galaxy image and the
artificial sources.  The results of this exercise, based on the {\sl
WFPC2} FoV, are shown in Fig.  \ref{compl.fig}.  These completeness
curves were corrected for the effects of blending or superposition of
multiple randomly placed artificial PSFs as well as for the
superposition of artificial PSFs on genuine objects. A detailed
description of the procedures employed to obtain these completeness
curves was given in de Grijs et al. (2002).

We found that the effects of image crowding are small: only $\lesssim
1.5$\% of the simulated objects were not retrieved by the {\sc daofind}
routine due to crowding.  However, the effects of the bright and
irregular background and dust lanes are large, resulting in variable
completeness fractions across the central galaxy images.  As a general
rule, however, the curves in Fig.  \ref{compl.fig} show that
incompleteness becomes severe (i.e.\ the completeness drops below $\sim
50$\%) for F300W $> 22.5$ mag, for F606W $> 24$ mag, and for F814W $>
24$ mag. 

Foreground stars are not a source of confusion (e.g., E02).  The
standard Milky Way star count models (e.g., Ratnatunga \& Bahcall 1985)
predict roughly 1--2 foreground stars for the equivalent standard filter
of F606W $\lesssim 24$ in the combined {\sl WFPC2} FoV of our images. 

Background objects may pose a (small) problem, however, in particular
among the fainter sources, since we did not impose any roundness or
sharpness constraints on our extended source detections, in order not to
omit unrelaxed young clusters from our final sample. However, such
objects should be easily identifiable once we have obtained aperture
photometry for our complete source lists, as they are expected to have
significantly different colours. More quantitatively, background
galaxies at redshifts greater than about 0.1 are expected to have
extremely red (F300W--F814W) colours compared to their local
counterparts. 

\section{Analysis of the star cluster photometry}
\label{photom.sec}

We applied a three-dimensional $\chi^2$ minimisation to the SEDs of our
star cluster candidates to obtain the most likely combination of age
{\it t}, metallicity {\it Z} and total (i.e., Galactic foreground and
internal) extinction E$(B-V)$ (assuming a Calzetti et al.  [2000]
starburst galaxy-type extinction law) for each individual object. 
Galactic foreground extinction is taken from Schlegel et al.  (1998) for
each individual object (Section \ref{agedet.sec}; see Anders et al. 
2002, 2003).  The least-squares minimisation of the clusters' SEDs was
done with respect to the G\"ottingen SSP models (Schulz et al.  2002),
to which we added the age and metallicity-dependent contributions of an
exhaustive set of gaseous emission lines and gaseous continuum emission
(Anders et al.  2002, Anders \& Fritze--v.  Alvensleben 2003).  Their
contributions to the broad-band fluxes are most important for
significantly subsolar abundances (i.e., $0.2 Z_\odot$; because the YSCs
{\it and} the ionised gas are assumed to have the same low metallicity)
during the first $\simeq 3 \times 10^7$ yr of their evolution (Anders et
al.  2002, Anders \& Fritze--v.  Alvensleben 2003), while for solar
metallicity populations the relative importance is significantly reduced
(to roughly half the relative contribution of $0.2 Z_\odot$ objects) and
of much shorter duration ($t \lesssim 1.2 \times 10^7$ yr). 

The Schulz et al.  (2002) G\"ottingen SSP models compose a model grid in
age and metallicity covering ranges of $0.02 \le Z/Z_\odot \le 2.5$ (as
discrete metallicities of $0.02, 0.2, 0.4, 1.0$ and $2.5 Z_\odot$) and
$1.4 \times 10^8 \le {\rm age/yr} \le 1.6 \times 10^{10}$ yr, in time
steps of $1.4 \times 10^8$ yr, respectively.  They are based on the most
recent Padova isochrones and include the thermally pulsing AGB stellar
evolutionary phase for stars with masses $m$ in the range $2 \le
m/M_\odot \le 7$.  We also extended the age range of the G\"ottingen SSP
models down to ages as young as $4 \times 10^6$ yr in time steps as
short as $4 \times 10^6$ yr for the youngest ages. 

In order to obtain useful fit results for all of our three free
parameters, i.e., age, metallicity and extinction\footnote{Strictly
speaking, the cluster mass is also a free parameter.  Our model SEDs are
valid for SSPs with masses of $1.6 \times 10^9 M_\odot$; to obtain the
actual cluster mass, we scale the model SED to match the observed
cluster SED using a single scale factor.  This scale factor is then
converted into a cluster mass.}, we need a minimum SED coverage of four
passbands (or three independent colours).  To assess the effects of
choosing a particular passband combination on the final results, we
started with applying our fitting technique to passband combinations
consisting of four filters each, and subsequently added more passbands,
as follows:

\begin{enumerate}

\item a blue-selected passband combination, consisting of photometry in
the F300W (``UV''), F336W ({\it ``U''}), F439W ({\it ``B''}) and F606W
({\it ``V''}) filters;

\item a subset of only optical passbands: F336W, F439W, F606W and F814W
({\it ``I''});

\item a red-selected passband combination: the red optical F606W and
F814W filters, combined with the NIR F110W ({\it ``J''}) and F160W
({\it ``H''}) passbands;

\item a five-passband optical/near-infrared combination: F439W,
F606W, F814W, F110W and F160W;

\item the previous combination with the addition of the blue F336W
filter; and

\item our full set of seven passbands from across the entire
wavelength range.

\end{enumerate}

Finally, in Section \ref{agemet.sec} we compare our results from the
G\"ottingen SSP models to independently determined solutions for the
same star cluster samples based on the Starburst99 SSP models (Leitherer
et al.  1999) for ages below 1 Gyr, and on the Bruzual \& Charlot (2000;
BC00) models for older ages.  The fitting of the observed cluster SEDs
to the Starburst99 and BC00 models was done using a three-dimensional
maximum likelihood method, ``3/2DEF'', with the initial mass $M_i$, age
and extinction E$(B-V)$ as free parameters (see Bik et al.  2003), and
assuming a fixed, solar metallicity, a Salpeter-type initial mass
function (IMF), and the (Galactic) extinction law of Scuderi et al. 
(1996).  We will discuss the uncertainties introduced by the assumption
of solar metallicity below.  For the clusters with upper limits in one
or more filters we used a two-dimensional maximum likelihood fit, using
the extinction probability distribution for E$(B-V)$.  This distribution
was derived for the clusters with well-defined SEDs over the full
wavelength range (see Bik et al.  2003). 

\section{Setting the scene: The young merger remnant NGC 3310}
\label{ngc3310.sect}

NGC 3310 is a local, very active starburst galaxy with high global star
formation rate, as is evident from its bright optical and infrared
continuum emission (Telesco \& Gatley 1984 [hereafter TG84], Braine et
al.  1993, Mulder, van Driel \& Braine 1995 [hereafter MvDB95], D\'\i az
et al.  2000 [hereafter D00], E02), strong and extended X-ray, UV, and
thermal radio emission (van der Kruit \& de Bruyn 1976, Code \& Welch
1982, Fabbiano, Feigelson \& Zamorani 1982, Zezas, Georgantopoulos \&
Ward 1998, Conselice et al.  2000), intense UV and optical emission
lines typical of OB associations (Heckman \& Balick 1980, Kinney et al. 
1993, Mulder et al.  1995 and references therein) and its large global
H$\alpha$ equivalent width (MvDB95, Mulder \& van Driel 1996 [hereafter
MvD96], and references therein). 

\subsection{A peculiar starburst galaxy}

A number of morphological peculiarities in its outer parts (e.g., Balick
\& Heckman 1981, Kregel \& Sancisi 2001, [hereafter BH81, KS01], MvD96),
combined with the disturbed kinematics of the H{\sc i} gas in the inner
regions (e.g., van der Kruit 1976, MvDB95, KS01), the mismatch between
stellar and gas-dynamical geometry (e.g., BH81), and the large number of
early-type stars required to explain the galaxy's H$\alpha$ emission
(van der Kruit 1976, BH81), suggest that NGC 3310 was affected by a
major gravitational disturbance.  This led to high, possibly sustained,
star formation rates in the past $\sim 100$ Myr (cf.  BH81).  Since
attempts to identify a nearby companion galaxy as the cause for the
disruption and the expected subsequent major starburst were unsuccessful
(cf.  van der Kruit 1976, van der Kruit \& de Bruyn 1976), it was
suggested that NGC 3310 accreted a gas-rich, but metal-poor companion
galaxy, which subsequently fragmented as a result of the encounter
(BH81, Schweizer \& Seitzer 1988, MvDB95, MvD96, Smith et al.  1996,
hereafter S96; modeled by Athanassoula 1992 and Piner, Stone \& Teuben
1995), or perhaps we are currently seeing a newly-formed disc.  This
argument was based predominantly on the absence of any close companion
galaxy and on the unusually low (subsolar) metallicity found in star
forming regions surrounding the nucleus, although the nucleus itself
appears to have solar metallicity (e.g., Heckman \& Balick 1980, Puxley,
Hawarden \& Mountain 1990, Pastoriza et al.  1993, hereafter P93). 
Additional support for this interpretation is provided by the
far-ultraviolet and {\it B}-band $R^{1/4}$ luminosity profiles, typical
of late-stage galaxy mergers (S96, KS01), and the optical and H{\sc i}
tails and ripples observed in the outer parts. 

The galaxy's most conspicuous visual peculiarity is the well-studied
bright optical ``bow and arrow'' structure (nomenclature first used by
Walker \& Chincarini 1967).  The ``arrow'' is a jet-like structure with
possibly an H{\sc i} counterpart (although the latter is significantly
more extended; MvDB95), which has been interpreted as the result of a
nuclear explosive event some $1.8 \times 10^7$ yr ago (Bertola \& Sharp
1984), or -- in combination with the ripple pattern that includes the
``bow'' -- as the result of a recent merger with a smaller companion
galaxy (or even a third, smaller object; KS01).  The latter explanation
seems more likely (see MvDB95, KS01).  These authors argued that a
``projectile'' object hitting the disc of NGC 3310 at a fairly oblique
angle could have been drawn out into the observed configuration, which
is supported by the anomalous velocity structure observed within these
features (MvDB95); the optical ``arrow'' is then best understood as a
series of regions in which active star formation was triggered due to
the collision.  If this scenario is correct, the merger event must have
occurred $\lesssim 10$ Myr ago to prevent significant redistribution of
the longer H{\sc i} jet due to differential rotation (cf.  MvDB95),
although the outer jet regions appear to curve away from the line of
sight (KS01, see also MvDB95). 

\subsection{Star cluster formation}

The bar-driven star formation scenario suggested by Conselice et al. 
(2000), combined with the recent infall of a companion galaxy, is very
attractive: it provides a natural explanation for the low metallicity
observed in the star-forming knots near the centre, while it also
explains why we observe concentrated star formation in star clusters or
very luminous H{\sc ii} regions in a tightly-wound ring-like structure
surrounding the centre (e.g., van der Kruit \& de Bruyn 1976, TG84, P93,
Meurer et al.  1995, S96, Conselice et al.  2000, E02), coinciding with
the end of the nuclear bar (Conselice et al.  2000, but see D00), but
not inside this ring. 

\begin{figure*}
\hspace*{1.2cm}
\psfig{figure=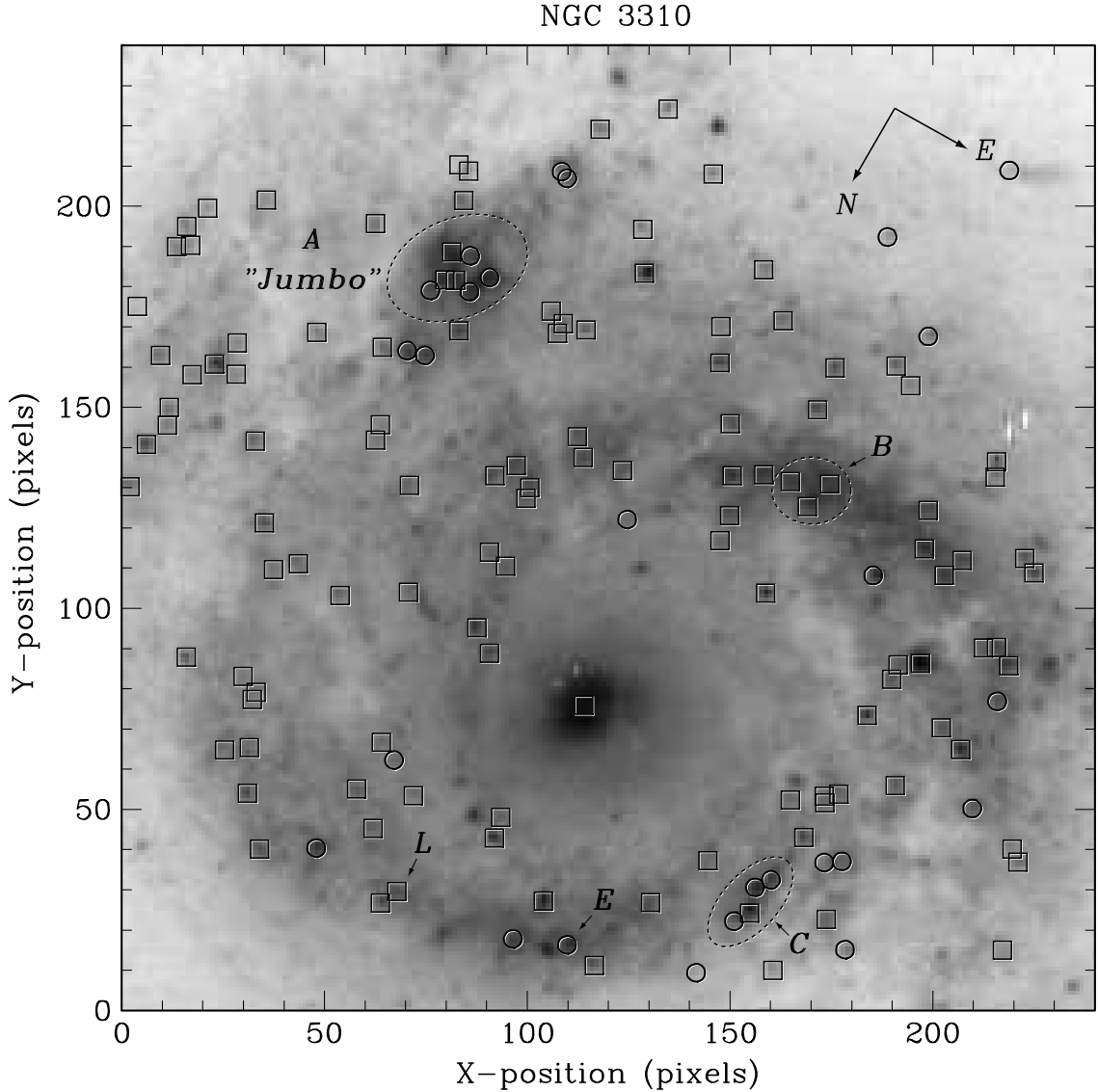,width=15cm}
\caption{\label{clustercoords.fig}Cluster detections in NGC 3310, for
which we obtained acceptable solutions for all of their ages,
metallicities, and extinction values (see Sect. \ref{coverage.sect})
overlaid on a grey-scale rendition of our F300W press release image of
the galaxy (Windhorst et al.  2001, 2002).  The circles represent
sources younger than $\log( {\rm age/yr} ) = 7.1$, the squares are the
older objects.  The nomenclature of sources A--C, E, and L is from P93.}
\end{figure*}

The most luminous single star-forming region in NGC 3310 is the giant
``Jumbo'' H{\sc ii} region (BH81, TG84, S96); it is roughly $10 \times$
more luminous than 30 Doradus in H$\alpha$ (BH81), but it is of very low
metallicity, $Z \sim 0.1 Z_\odot$ (P93).  The Jumbo H{\sc ii} region
contains several individual UV-bright star clusters (Meurer et al. 
1995, E02).  The most luminous star-forming structure is the ring of
H{\sc ii} regions, which produces some 30\% of the observed far-UV flux
of NGC 3310 (S96).  We have indicated our cluster detections in Fig. 
\ref{clustercoords.fig}, overlaid on our press release image of the
galaxy (Windhorst et al.  2001). 

Numerous other, less luminous star clusters and H{\sc ii} regions, with
sizes from 10 pc for the most luminous ones to $\lesssim 1$ pc for the
unresolved clusters in the background disc (see Conselice et al.  2000),
are found scattered in the galaxy's outer parts beyond $R \simeq 4$ kpc
(van der Kruit \& de Bruyn 1976, BH81, TG84, MvDB95, S96) and as
condensations in the ``bow and arrow'' structure (e.g., Bertola \& Sharp
1984, S96, KS01); these clusters may have been produced by the accretion
event, or might be remnants of the progenitor companion galaxy (cf. 
BH81, S96). 

\subsection{Star formation time-scales}

All of the observational evidence points at very recent star formation
in the star clusters and H{\sc ii} regions, and a time since the
interaction of $\lesssim 10^7-10^8$ yr (van der Kruit 1976, BH81, TG84,
P93, S96, E02): the optical colours and the equivalent widths of the
H$\alpha$-bright circumnuclear sources are best reproduced by a
combination of a 2.5 Myr and an 8 Myr-old population (cf.  P93, D00). 
This is consistent with {\it (i)} the detection of WR features in the
spectra of a few of the H{\sc ii} regions, including the Jumbo region,
and of the NIR Ca II triplet (Terlevich et al.  1990, P93), both
indicative of stellar populations with ages $\lesssim 4$ Myr, and {\it
(ii)} the absence of significant non-thermal radio emission (cf.  TG84),
which implies that there is not yet a significant population of
supernova remnants.  Terlevich et al.  (1990) found evidence for two
stellar populations (of 5 and 15 Myr old) as well, which can also be
interpreted as an extended star formation episode.  As pointed out by
TG84 (see also MvDB95), the star formation in NGC 3310 can continue at
its present rate only for a small fraction of the age of the galaxy,
roughly $4 \times 10^7 - 10^8$ yr, depending on the IMF assumed
(MvDB95). 

\section{Age determinations: systematic uncertainties}
\label{agedet.sec}

\subsection{Dependence on the SED coverage}
\label{coverage.sect}

We obtained 156, 145, 100, 85, 63 and 60 solutions from our
least-squares minimisation technique (for our age, metallicity and
extinction determinations), respectively, for the six passband
combinations chosen for the analysis of the NGC 3310 photometry (see
Sect.  \ref{photom.sec}) in the small FoV covered by the full passband
combination.  The relatively small fraction of acceptable solutions
mainly reflects the rather low S/N ratios in a number of our archival
images, resulting in either large photometric uncertainties or upper
limits for the aperture photometry.  Since the main aim of this paper is
to analyse the dependence on the choice of passband combination of the
resulting ages, extinction values and metallicities, objects containing
upper limits to their photometry in any of the passbands in a given
passband combination were excluded from the fitting procedure.  We
emphasize once again that we used exactly the same sample of star
clusters, with well-determined photometric measurements in {\it all}
passbands (as defined above), for this comparison.  Acceptable solutions
were required to have fit values for {\it each} photometric measurement
in a given passband combination within the $2\sigma$ observational
uncertainty associated with that particular filter; weights were
allocated proportional to $1/\chi^2$. 

\begin{figure*}
\hspace{1.2cm}
\psfig{figure=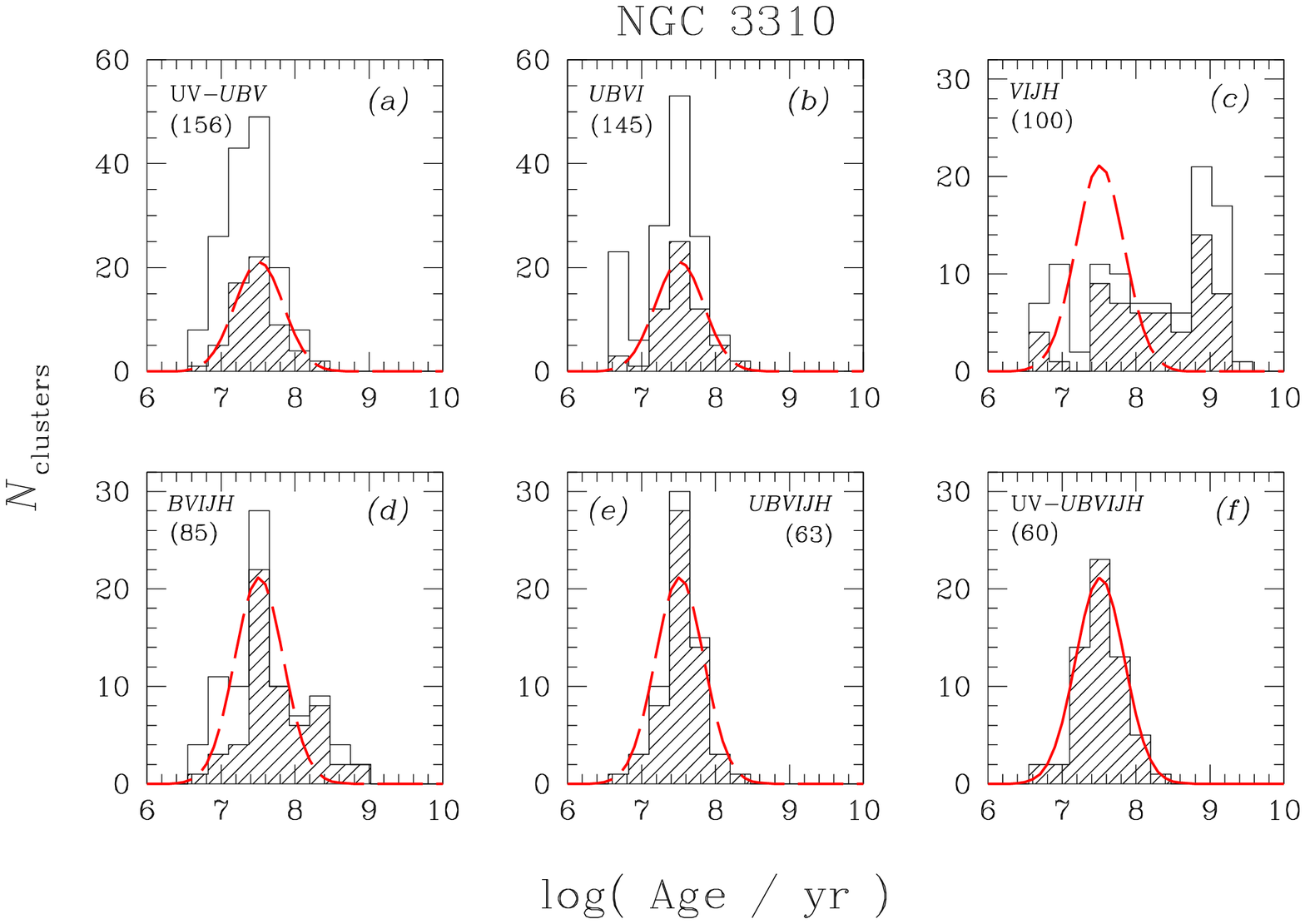,width=15cm}
\vspace{-4.5cm}
\caption{\label{n3310ages.fig}Relative age distributions of the NGC 3310
clusters based on six choices of broad-band passband combinations, with
acceptable solutions.  The shaded histograms represent the cluster
sample with acceptable solutions in common among all six passband
combinations (60 clusters); the open histograms are the additional
solutions obtained for subsamples based on the passband combination
displayed in each panel.  The numbers in between brackets in each panel
correspond to the total number of clusters in the histograms.  We have
overplotted the best-fit Gaussian age distribution obtained in panel f.}
\end{figure*}

Fig.  \ref{n3310ages.fig} shows the dependence of the resulting age
distribution on the choice of passband combination.  The shaded
histograms represent the acceptable age solutions (with minimum $\chi^2$
values obtained from the least-squares minimisation) for the cluster
sample in common among all six passband combinations; the open
histograms are the additional solutions obtained for the respective
passband combinations.  The uncertainties in the {\it number} of
clusters in a given age bin are predominantly Poissonian; the
uncertainties in the actual ages are the subject of our current
analysis.  The electronic data tables containing our best solutions for
all passband combinations are also available from the aforementioned WWW
address.  It is immediately clear from this figure that the resulting
age distribution of an extragalactic star cluster system based on
broad-band colours is a sensitive function of the passbands covered by
the observations. 

To guide the eye, we have overplotted -- on all panels -- the best-fit
Gaussian age distribution based on the full, 7-passband SED sampling
(Fig.  \ref{n3310ages.fig}f).  This best-fit age distribution is
characterized by a peak at $\log( {\rm age / yr} ) = 7.51 \pm 0.12$ and
a Gaussian width of $\sigma_{\rm Gauss} = 0.33$, irrespective of the bin
size adopted. 

From a comparison of the age distributions in Fig. \ref{n3310ages.fig},
we determine that:

\begin{itemize}

\item the peak of the age distribution is robustly reproduced in all
cases, for both the shaded and the full cluster samples with acceptable
solutions, except for the red-selected passband combination (Fig.
\ref{n3310ages.fig}c); 

\item photometric measurements in red-dominated passband combinations
(Figs.  \ref{n3310ages.fig}c and d) result in significantly older (but
highly uncertain) age solutions.  This is due to the weak time
dependence of the NIR magnitudes and due to unavoidable ambiguities in
the modelling of the thermally pulsing AGB phase, thus resulting in
badly constrained ``best'' fit ages.  Red-dominated age solutions
produce a significant wing of older clusters, compared to the ages
derived from the full passband combination. 

\item blue-selected passband combinations (such as in Figs. 
\ref{n3310ages.fig}a and b; open histograms) tend to result in age
estimates that are slightly skewed towards younger ages, compared to
passband combinations that also include redder passbands.  This is
simply due to the combination of observational selection effects, in the
sense that younger clusters have higher UV and {\it U}-band fluxes and
are therefore more easily detected at those blue wavelenghts, and to the
age--metallicity degeneracy (see below).  Such secondary peaks at young
ages should therefore be treated with extreme caution. 

\item the optical passband combination (Fig.  \ref{n3310ages.fig}b), in
particular, displays a distinct secondary peak at very young ages,
$\log( {\rm age/yr} ) \sim 6.7$ (cf.  open histogram).  These clusters
do not appear to be exceptional in their metallicity, extinction or mass
properties compared to the overall distribution of these properties
among the other clusters in our sample.  However, with few exceptions,
we could not obtain satisfactory solutions for these clusters as part of
our full passband combination fits.  As we will show below, this
secondary peak is most likely the result of the age--metallicity
degeneracy. 

\end{itemize}

\begin{figure}
\psfig{figure=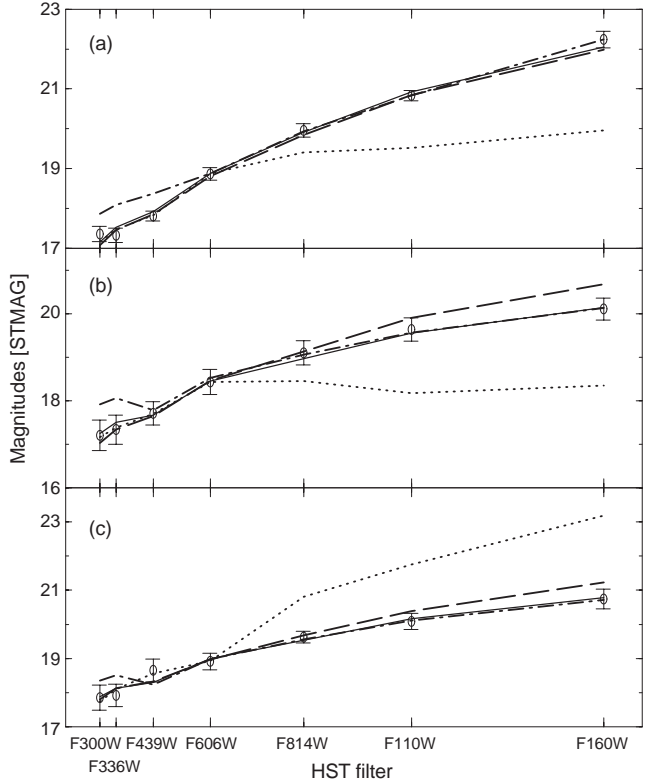,width=8.5cm}
\caption{\label{indivclus.fig}Illustration of the sensitivity of the
parameter determinations on the available passband combination for three
representative clusters in regions ``C'' (panel a) and ``B'' (panels b
and c).  The dotted lines represent the best-fitting models based on a
blue-selected passband combination, the dash-dotted lines were derived
from the red-selected combination, while the dashed and solid fits are
based on the full passband combination and on the subset of $UBVI$
filters only, respectively. The {\sl HST} filters are spaced according
to their central wavelengths.}
\end{figure}

In Fig.  \ref{indivclus.fig} we illustrate the sensitivity of the
parameter determinations on the available passband combination by
showing the SEDs and best fits for three typical, representative
clusters, located in P93 regions ``C'' (Fig.  \ref{indivclus.fig}a;
cluster No.  18) and ``B'' (Figs.  \ref{indivclus.fig}b and c; Nos.  116
and 122, respectively).  The dotted lines represent the best-fitting
models based on a blue-selected passband combination, the dash-dotted
lines were derived from the red-selected combination, while the
dashed and solid fits are based on the full passband combination
and on the subset of $UBVI$ filters only, respectively.  In Table
\ref{indivclus.tab} we list the best-fitting free parameters as a
function of passband combination used for all three clusters, to
illustrate the effects of our choice of wavelength coverage. 

\begin{table}
\caption[ ]{\label{indivclus.tab}Dependence of our fit parameters on the
available passband combination for three representative clusters
}
{\scriptsize
\begin{center}
\begin{tabular}{lcccc}
\hline
\hline
\multicolumn{1}{c}{Passbands} & \multicolumn{1}{c}{log( Age )} &
\multicolumn{1}{c}{log( Mass )} & \multicolumn{1}{c}{Metallicity} &
\multicolumn{1}{c}{E$(B-V)$} \\
 & \multicolumn{1}{c}{[yr]} &
\multicolumn{1}{c}{[$M_\odot$]} & \multicolumn{1}{c}{$(Z)$} &
\multicolumn{1}{c}{(mag)} \\
\hline
\multicolumn{5}{c}{Cluster No. 18 (from Region ``C'')}\\
UV$-UBV$    & 7.205 & 5.666 & 0.022 & 0.05 \\
$UBVI$      & 6.876 & 5.356 & 0.008 & 0.00 \\
$VIJH$      & 6.660 & 5.304 & 0.013 & 0.25 \\
UV$-UBVIJH$ & 7.204 & 5.512 & 0.002 & 0.00 \\
\\
\multicolumn{5}{c}{Cluster No. 116 (from Region ``B'')}\\
UV$-UBV$    & 7.362 & 5.968 & 0.011 & 0.08 \\
$UBVI$      & 7.448 & 5.939 & 0.008 & 0.06 \\
$VIJH$      & 8.939 & 6.770 & 0.002 & 0.06 \\
UV$-UBVIJH$ & 7.478 & 5.911 & 0.012 & 0.06 \\
\\
\multicolumn{5}{c}{Cluster No. 122 (from Region ``B'')}\\
UV$-UBV$    & 6.986 & 5.474 & 0.012 & 0.13 \\
$UBVI$      & 7.276 & 5.577 & 0.010 & 0.10 \\
$VIJH$      & 8.071 & 6.079 & 0.003 & 0.12 \\
UV$-UBVIJH$ & 7.349 & 5.636 & 0.012 & 0.11 \\
\hline
\end{tabular}
\end{center}
}
\end{table}

Note that all of our conclusions apply to the YSC population in NGC
3310, characterised by ages in the range from $\sim 10^6$ to $\sim 10^9$
yr.  The situation may be significantly different for clusters of
significantly greater age.  Thus, from an observational point of view,
for ages $6 \lesssim \log( {\rm age/yr} ) \lesssim 9$, we conclude that
if one can only obtain partial coverage of a star cluster's SED, an
optical passband combination including {\it both} blue {\it and} red
optical passbands results in the best balanced and most representative
age distribution, as compared to the better constrained ages obtained
from the full UV--NIR SED coverage.  While blue-selected passband
combinations lead to age distributions that are slightly biased towards
younger ages, red-selected passband combinations -- in particular if
they are dominated by NIR filters -- should clearly be avoided. 

\subsection{The age--metallicity degeneracy}
\label{agemet.sec}

\begin{figure*}
\hspace{1.2cm}
\psfig{figure=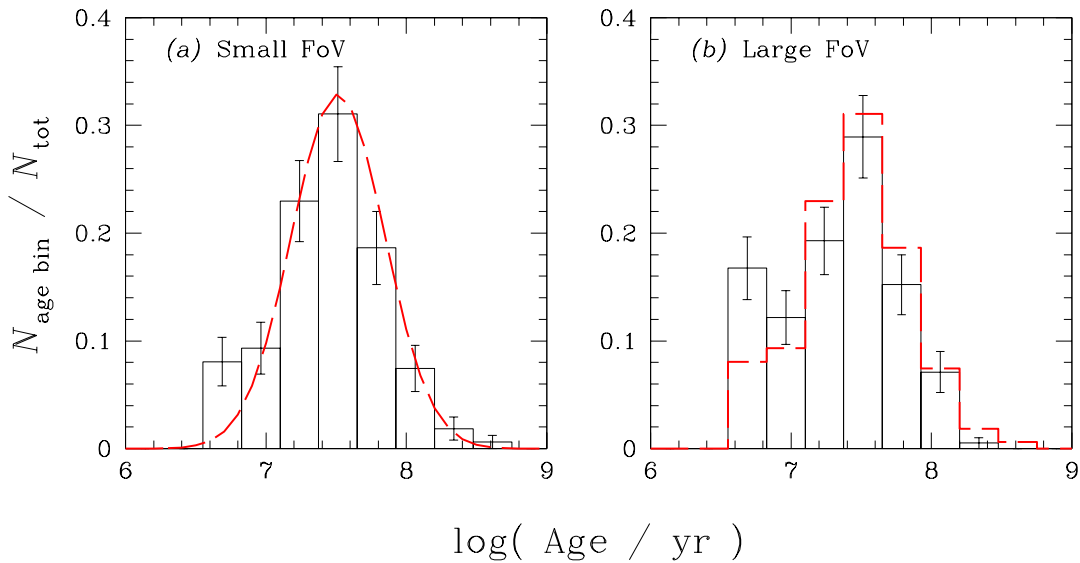,width=15cm}
\vspace{-7cm}
\caption{\label{cffields.fig}Comparison of the age distributions of the
NGC 3310 star clusters in the small, inner FoV and the large, {\sl
WFPC2}-only FoV.  The error bars represent Poissonian uncertainties in
the numbers of clusters in each age bin.  The dashed Gaussian
distribution (panel a) is our best-fit age distribution (Fig. 
\ref{n3310ages.fig}f); the dashed histogram overplotted on the large FoV
age distribution (panel b) is the age distribution of the small FoV.}
\end{figure*}

Finally, in Fig.  \ref{cffields.fig} we compare the best-fit age
distributions of the clusters in the small FoV covered by all seven
passbands, and the larger ``{\sl WFPC2}-only'' FoV, covered by our five
optical passbands.  These best-fit age distributions were derived from
either the full, seven/five passband SED coverage if acceptable
solutions could be obtained for all of our free parameters, or from the
passband combination with the highest ``confidence ranking'', in a
minimum $\chi^2$ sense: we defined a confidence ranking among the
passband combinations, such that a higher ranking was given to solutions
obtained from a passband combination (i) containing more passbands, (ii)
covering a larger wavelength base line and (iii) resulting in a smaller
$\chi^2$ value. 

For comparison, in Fig.  \ref{cffields.fig}a we have -- once more --
overplotted the best-fit Gaussian age distribution obtained from the
full SED sampling of the same (small) FoV, as in Fig. 
\ref{n3310ages.fig}f.  The vertical error bars represent the Poissonian
uncertainties dominating the cluster numbers in each age bin. In Fig. 
\ref{cffields.fig}b we show both the age distribution of the clusters in
the large FoV (solid histogram) and again the age distribution in the
small FoV, overplotted as the dashed histogram (both normalised by the
total number of clusters in each sample), in order to illustrate the
apparent $\sim (2-2.5) \sigma$ peak at young ages, $\log( {\rm age/yr} )
\lesssim 6.8$. 

A naive interpretation of this feature would suggest that the outer
field of NGC 3310, outside the small FoV, contains a subpopulation of
significantly younger clusters.  However, closer inspection of our data
reveals that this young peak is entirely owing to unavoidable systematic
uncertainties inherent to our fitting technique and the shape of the
SED.  It turns out that the secondary, young peak is caused by the
limited number of passbands available for the full cluster sample in the
{\sl WFPC2}-only FoV, due to the lack of NIR photometry for the clusters
outside the small FoV.  If we restrict our fits of the age, extinction
and metallicity properties of the cluster sample in the {\it small} FoV
to the same {\sl WFPC2}-only passbands, we find a similar young age peak
in the inner FoV, of similar strength and significance. 

The clusters contained in this latter young-age peak are characterized
by a smoother age distribution (ranging from $\sim 5 \times 10^6$ to
$\sim 2 \times 10^7$ yr, with a few outliers as old as $\sim 10^8$ yr)
in the best-fit age distribution based on the full UV--NIR SED sampling. 
In addition, while these clusters are all characterized by a narrow
spread in metallicity around $\sim 0.5 Z_\odot$ in the restricted fit
using the {\sl WFPC2} passbands only, their metallicity estimates range
from $\sim 0.02 Z_\odot$ to $\sim 2.5 Z_\odot$ if we take the full
UV--NIR SED into account.  Thus, we conclude that the secondary age peak
at young ages in the {\sl WFPC2}-only FoV is an artefact caused by the
age--metallicity degeneracy at ages younger than about $10^8$ yr.  The
age--extinction degeneracy is not as significant as this
age--metallicity degeneracy, for these particular clusters.  The slight,
$\sim 2 \sigma$ excess of young clusters in the youngest age bin of Fig. 
\ref{cffields.fig}a may be a remaining systematic effect, or may indeed
be marginal evidence of more recent cluster formation. 

\begin{figure}
\psfig{figure=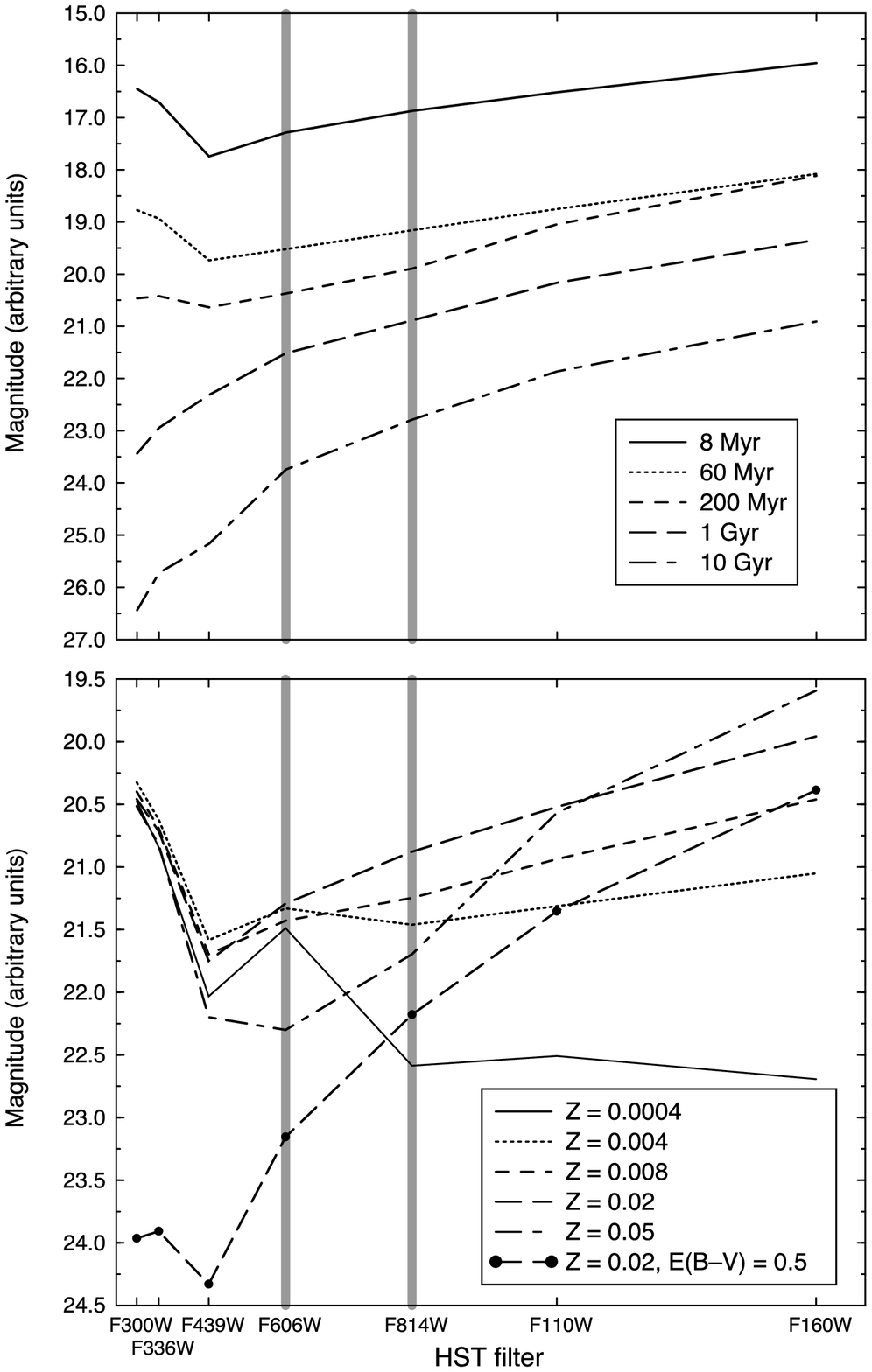,width=8.5cm}
\caption{\label{models.fig} {\it (top panel)} Age dependence of the
model SEDs for solar metallicity.  {\it (bottom panel)} Metallicity
dependence of the 8 Myr-old model SED.  we also show the effects of
adding E$(B-V) = 0.5$ mag extinction to the solar-metallicity SED
(dashed line with solid bullets).  The {\sl HST} filters are spaced
according to their central wavelengths.  The grey vertical lines are
drawn at the central wavelengths of the F606W (or F555W) and F814W {\sl
HST} filters often used for age dating of individual clusters.}
\end{figure}

To illustrate this point, in the top panel of Fig.  \ref{models.fig} we
show the age dependence of the model SEDs for solar metallicity.  Next,
in the bottom panel we explore the effects of varying the metallicity
for the youngest, 8 Myr-old SED.  For illustrative purposes we also show
the effects of adding E$(B-V) = 0.5$ mag extinction to the
solar-metallicity SED (dashed line with solid bullets).  The grey
vertical lines are drawn at the central wavelengths of the F606W (or
F555W) and F814W {\sl HST} filters often used for age dating of
individual clusters.  It is clear that for these young ages, the effects
of varying the metallicity, extinction or age of the stellar population
cannot be disentangled unless one has access to observations in a larger
number of passbands, which may allow to more robustly constrain all of
the free fitting parameters. 

\begin{figure}
\psfig{figure=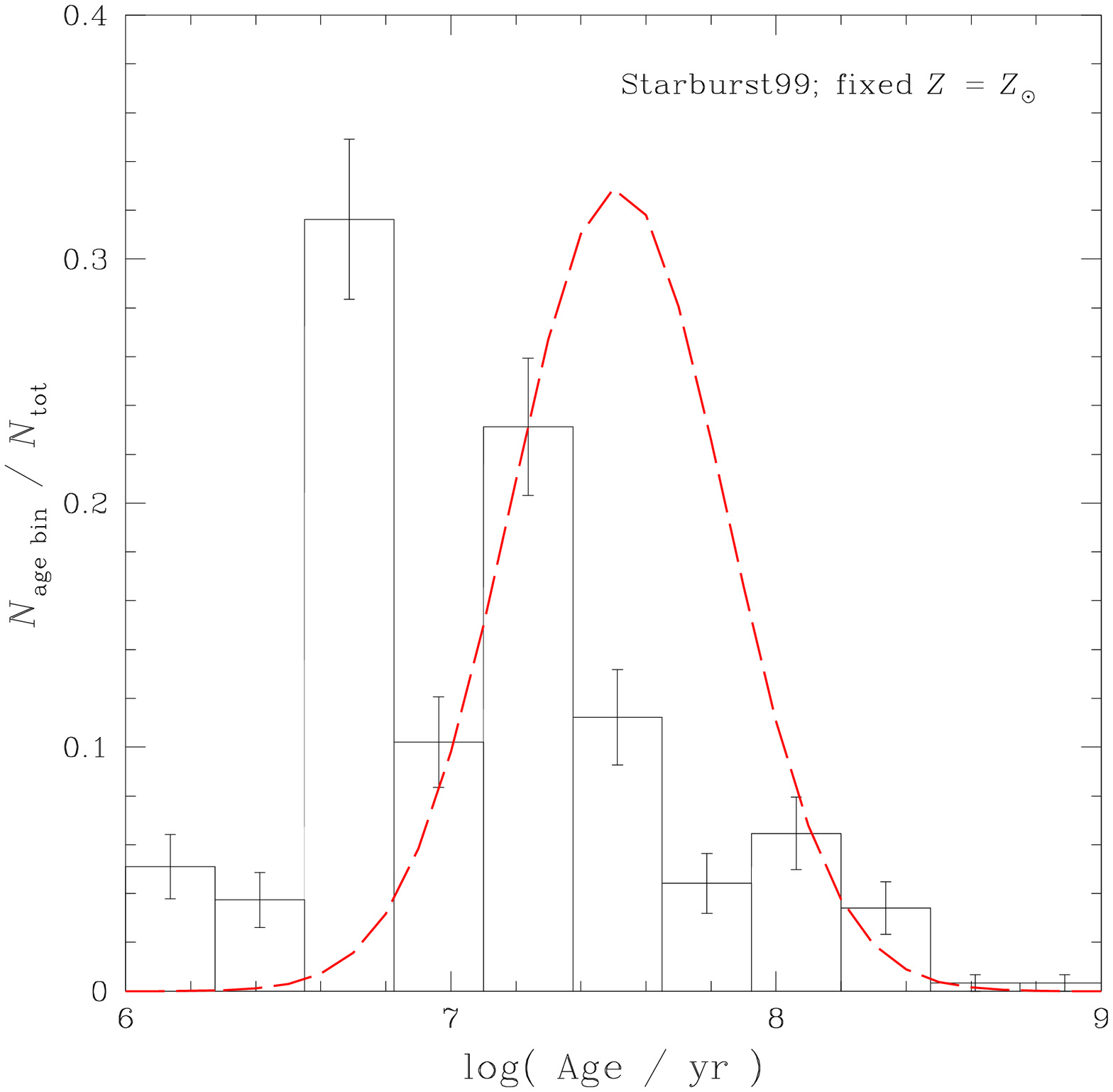,width=8.5cm}
\caption{\label{sb99fits.fig}Age distribution of the NGC 3310 star
clusters based on a comparison of their SEDs with the Starburst99 models
and fixed, solar metallicity.  The dashed Gaussian distribution is our
best-fit age distribution from a comparison with the G\"ottingen SSP
models, leaving the metallicity as a free parameter.  The error bars
represent Poissonian uncertainties in the numbers of clusters in each
age bin.}
\end{figure}

Finally, in Fig.  \ref{sb99fits.fig} we compare our best fit age
distribution (dashed Gaussian distribution) with the best fit age
distribution using the Bik et al.  (2003) approach.  The latter is based
on the Starburst99/BC00 SSP models, and a fixed metallicity of $Z =
Z_\odot$.  While the differences between the age distributions resulting
from fitting the G\"ottingen or the Starburst99/BC00 models are expected
to be random due to the use of different stellar evolutionary tracks and
different template spectra for (super)giant stars (see Schulz et al. 
2002), the assumption of fixed, solar metallicity causes the fitting
technique to produce a significantly different age distribution compared
to the distribution obtained by leaving the metallicity as a free
parameter.  We confirmed this result by fitting the individual cluster
ages and extinction values using the G\"ottingen SSP models, while
keeping the metallicity fixed at solar values.  This is, again, a clear
signature of the age--metallicity degeneracy.  We point out that this
is, in fact, an expected effect, since unusually low (subsolar)
metallicities were found in NGC 3310's star forming regions (e.g.,
Heckman \& Balick 1980, Puxley, Hawarden \& Mountain 1990, P93, D00; $Z
\sim (0.2 - 0.4) Z_\odot$), so that the assumption of solar metallicity
for these sources must have a significant systematic effect. 

\subsection{Metallicity and extinction estimates}

\begin{figure*}
\hspace{1.2cm}
\psfig{figure=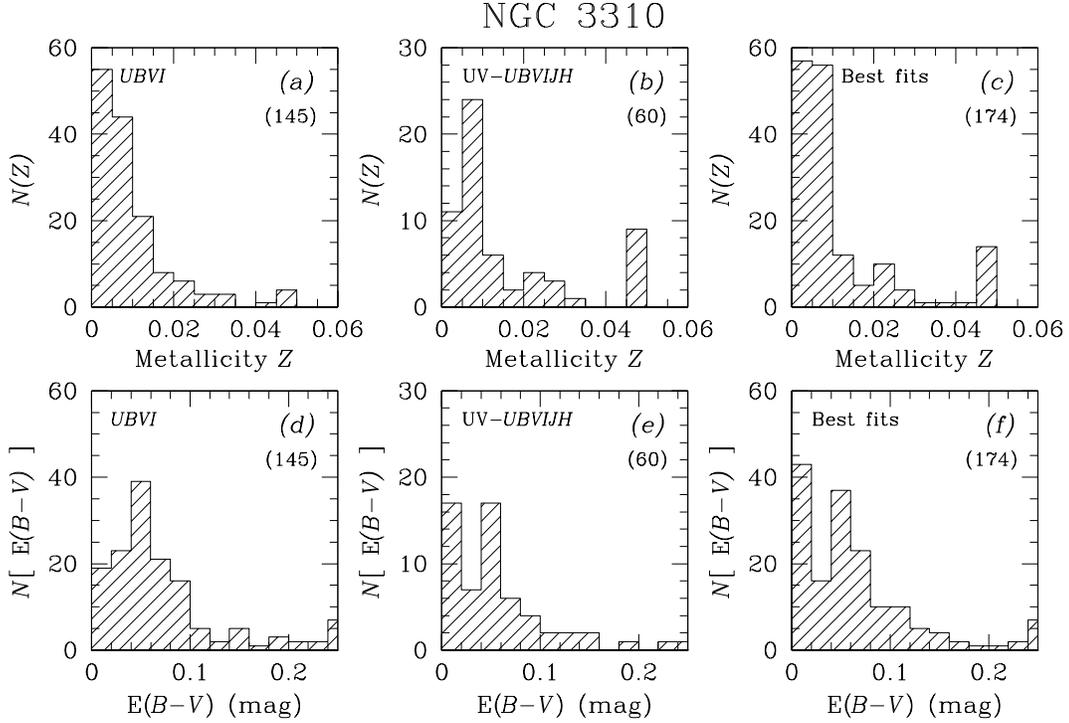,width=15cm}
\vspace{-5cm}
\caption{\label{met3310.fig}Distributions of metallicity, and of total
extinction towards the NGC 3310 clusters for the solutions from a number
of passband combinations.  The numbers in between parentheses correspond
to the total number of clusters represented by each histogram.}
\end{figure*}

Figure \ref{met3310.fig} shows our best estimates of the global
distribution of metallicity, and of total extinction (i.e., Galactic
foreground and internal extinction) towards the NGC 3310 clusters.  At
first sight, the metallicity and extinction distributions obtained from
either of the passband combinations used for Fig.  \ref{met3310.fig}
appear very similar within the (systematic) uncertainties.  The effects
of the age--metallicity degeneracy for the optical passbands (Fig. 
\ref{met3310.fig}a) are clearly visible.  Based on the {\it UBVI}
passband combination only, we find that the metallicity distribution of
the clusters in NGC 3310 is strongly dominated by significantly subsolar
metallicities, while the addition of NIR passbands results in the
detection of a significant number of clusters with metallicities $Z \sim
2.5 Z_\odot$.  In either case, the fact that the resulting metallicity
determinations are strongly dominated by (significantly) subsolar
metallicities is encouraging, in view of independent metallicity
measurements in the literature.  Both the low and higher-metallicity
objects are distributed fairly smoothly across the face of the galaxy;
there is some evidence that the most actively star forming regions, in
particular the Jumbo region and the northern spiral arm (including
complex C), are predominantly composed of lower-abundance star clusters. 

Similarly, the addition of NIR passbands appears to result in more
solutions with lower extinction estimates (Figs.  \ref{met3310.fig}e and
f).  This is likely due to the better constraints on the extinction from
optical-NIR compared to optical-only baselines.  However, whether this
result is indeed significant also depends on the suitability of the
starburst galaxy-type extinction law adopted for our fitting routine
(Calzetti et al.  2000), of which the analysis is beyond the scope of
the present paper (but see Calzetti, Kinney, \& Storchi-Bergmann [1994]
for a discussion).  Nevertheless, these effects are negligible, or
marginally significant at worst, compared to the photometric
uncertainties ($\sim 0.10-0.15$ mag). 

\section{NGC 3310 in its physical context}
\label{interpretation.sec}

\subsection{Overall properties of the NGC 3310 cluster population}

We can conclude from this exercise that NGC 3310 underwent a significant
burst of cluster formation some $3 \times 10^7$ yr ago.  The actual
duration of the burst of cluster formation may have been shorter because
uncertainties in the age determinations may have broadened the peak.  It
appears, therefore, that the peak of cluster formation in NGC 3310
coincides closely with the suspected galactic cannibalism or last tidal
interaction (van der Kruit 1976, BH81, TG84, P93, S96, E02), while the
possibly more recent, marginally significant cluster formation at $t
\lesssim 10^7$ yr can be interpreted as cluster formation associated
with the most recent ($\lesssim 10$ Myr) enhanced star formation episode
traced by the circumnuclear H$\alpha$-bright sources (P93, D00). 

The clusters older than $\sim 10^7$ yr are smoothly distributed
throughout the galactic centre, roughly following the inner ring
structure and the other concentrations of star clusters (see Fig. 
\ref{clustercoords.fig}).  However, the younger clusters, with ages
$\log( {\rm age/yr}) \le 7.1$ (i.e., the youngest two age bins in Fig. 
\ref{n3310ages.fig}), are predominantly concentrated in the Jumbo region
(``A''; nomenclature from P93) and in the northern spiral arm, which
also contains the bright star-forming regions C and E (see Fig. 
\ref{clustercoords.fig}).  This is consistent with the very young ages
derived for these regions, $t \lesssim 10$ Myr (Terlevich et al.  1990,
P93, D00, E02), although we emphasize that both regions also contain
clusters spanning the entire age range observed for the NGC 3310 star
cluster system. 

\begin{figure*}
\hspace{1.2cm}
\psfig{figure=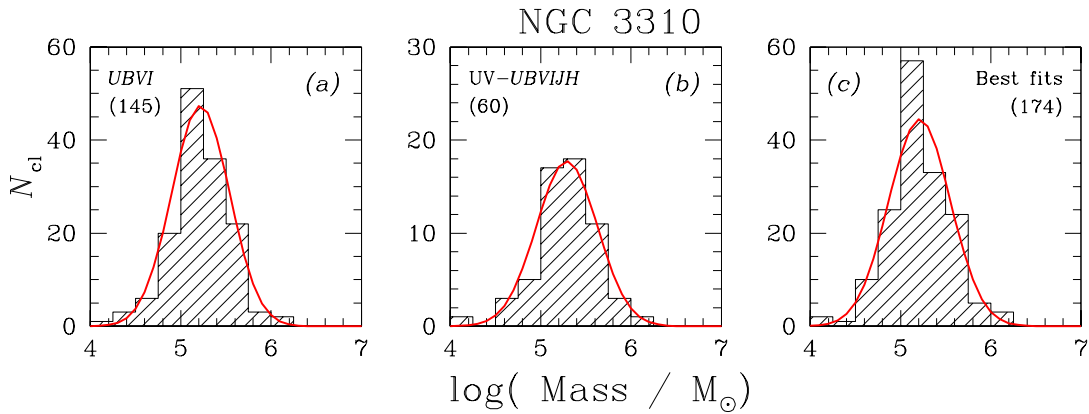,width=15cm}
\vspace{-9cm}
\caption{\label{mass3310.fig}Mass distributions for the NGC 3310
clusters for a number of passband combinations. Overplotted are the
best-fit Gaussian distributions (solid lines). The numbers in between
brackets correspond to the total number of clusters represented by each
histogram.}
\end{figure*}

Figure \ref{mass3310.fig} shows our best estimates of the clusters'
masses.  These were obtained by scaling our model SEDs (for masses of
$1.6 \times 10^9 M_\odot$) to the observed SEDs for the appropriate
combination of age, metallicity and (total) extinction.  Our model SEDs
are based on a Salpeter-type IMF consisting of stellar masses {\it m} in
the range $0.15 \le m/M_\odot \le (50-70)$ (Schulz et al.  2002).  The
exact upper mass limit for our model SSPs depends on the metallicity and
is determined by the mass coverage of the Padova isochrones.  The exact
value for the upper mass limit is unimportant for the determination of
the total cluster masses. 

We realise that recent determinations of the stellar IMF deviate
significantly from a Salpeter-type IMF at low masses, in the sense that
the low-mass stellar IMF may well be significantly flatter than the
Salpeter slope.  The implication of using a Salpeter-type IMF for our
cluster mass determinations is therefore that we may have {\it
overestimated} the individual cluster masses (although the relative mass
distribution of our entire cluster sample remains unaffected). 
Therefore, we used the more modern IMF parametrisation of Kroupa, Tout
\& Gilmore (1993, hereafter KTG) to determine the correction factor,
$C$, between our masses and the masses obtained from the KTG IMF (both
normalised at $1.0 M_\odot$).  This IMF is characterised by slopes of
$\alpha = -2.7$ for $m > 1.0 M_\odot$, $\alpha = -2.2$ for $0.5 \le
m/M_\odot \le 1.0$, and $-1.85 < \alpha < -0.70$ for $0.08 < m/M_\odot
\le 0.5$.  Depending on the adopted slope for the lowest mass range, we
have therefore overestimated our individual cluster masses by a factor
of $1.5 < C < 2.4$ for an IMF containing stellar masses in the range
$0.15 \le m/M_\odot \lesssim 70$. 

The mass distributions derived from the different passband combinations
are fairly similar.  We determined the defining parameters of the
cluster mass distributions by fitting Gaussian distributions to them. 
The small differences among the median mass [$\langle \log(m/M_\odot)
\rangle = 5.22, 5.29$, and 5.21 for Figs.  \ref{mass3310.fig}a, b and c,
respectively] and Gaussian widths of the distributions (similarly
varying among $\sigma_{\rm Gauss} = 0.32, 0.34$, and 0.34, respectively)
represent the minimum systematic uncertainties in these parameter
determinations (minimum because they are based on our highest-confidence
passband combinations).  In view of the uncertainties introduced by the
badly known lower-mass slope of the adopted IMF, the median mass
estimates of the NGC 3310 cluster system are remarkably close to those
of the Galactic GC system and of the intermediate-age cluster system in
M82 B (de Grijs et al.  2003a,b). 

\subsection{Comparison of individual cluster results with previous
determinations}

Although both NGC 3310 as such and its star cluster system in particular
have been studied extensively, determinations of individual cluster
ages, masses, metallicities and extinction values are rare.  The most
extensive cluster sample useful for a comparison to the results
presented in this paper was published by E02.  They estimate masses and
ages of (a subset of) 11 large-scale, diffuse cluster complexes (see
also D00) and of 17 super star cluster (SSC) candidates based on {\sl
HST} photometry; the latter were also noted by Conselice et al.  (2000). 
In Table \ref{comparison.tab} we compare the E02 mass and age
determinations for their sample of SSC candidates, based on the
assumption of a mean metallicity of $Z = 0.008 = 0.4 Z_\odot$, with the
corresponding parameters derived in this paper. 

We give our mass estimates for two assumptions of the lower mass limit
of the Salpeter-type IMF; varying the upper mass limit has very little
effect.  While we have used a lower mass limit of $0.15 M_\odot$
throughout this paper (resulting in the mass estimates $M_{\rm tot,1}$),
E02's cluster mass estimates ($M_{\rm tot,cf}$) are based on the
Starburst99 models, which assume a lower mass cut-off of $1 M_\odot$. 
Everything else being equal, our resulting mass estimates are therefore
expected to be a factor of $\sim 13$ higher compared to theirs.  Our
total mass estimates assuming a lower mass cut-off of $1 M_\odot$ are
referred to as $M_{\rm tot,2}$. 

\begin{table*}
\caption[ ]{\label{comparison.tab}Comparison of our derived cluster
parameters with those of the E02 SSC candidates.
}
{\scriptsize
\begin{center}
\begin{tabular}{lccccrrc}
\hline
\hline
\multicolumn{1}{c}{Number} & \multicolumn{1}{c}{log( Age )} &
\multicolumn{1}{c}{log( Mass )} & \multicolumn{1}{c}{Metallicity} &
\multicolumn{1}{c}{E$(B-V)$} & \multicolumn{1}{c}{$M_{\rm tot,1}$} &
\multicolumn{1}{c}{$M_{\rm tot,2}$} & \multicolumn{1}{c}{$M_{\rm
tot,cf}$}
\\
\multicolumn{1}{c}{(E02)} & \multicolumn{1}{c}{[yr]} &
\multicolumn{1}{c}{[$M_\odot$]} & \multicolumn{1}{c}{$(Z)$} &
\multicolumn{1}{c}{(mag)} & \multicolumn{1}{c}{$(10^5 M_\odot)$} &
\multicolumn{1}{c}{$(10^4 M_\odot)$} & \multicolumn{1}{c}{$(10^4
M_\odot)$} \\
\hline
8        & 7.412  & 5.529 & 0.009  & 0.05 &  3.4 &  2.6 &  9 \\
19       & 7.331  & 6.175 & 0.001  & 0.00 & 15.0 & 11.5 & 45 \\
32       & 8.182  & 5.722 & 0.050  & 0.00 &  5.3 &  4.1 &  8 \\
36       & 6.989  & 5.629 & 0.001  & 0.09 &  4.3 &  3.3 &  6 \\
37       & 7.204  & 5.512 & 0.001  & 0.00 &  3.3 &  2.5 &  7 \\
49       & 7.973  & 6.006 & 0.008  & 0.15 & 10.1 &  7.8 &  8 \\
59       & 8.505  & 5.935 & 0.012  & 0.12 &  8.6 &  6.6 &  2 \\
64       & 7.512  & 5.077 & 0.050  & 0.00 &  1.2 &  0.9 &  2 \\
80       & 7.369  & 5.133 & 0.028  & 0.00 &  1.4 &  1.0 &  2 \\
81       & 7.136  & 4.832 & 0.008  & 0.05 &  0.7 &  0.5 &  2 \\
85       & 7.982  & 6.021 & 0.008  & 0.05 & 10.5 &  8.1 &  8 \\
90       & 7.139  & 4.935 & 0.050  & 0.00 &  0.9 &  0.7 &  3 \\
107      & 8.035  & 5.521 & 0.012  & 0.03 &  3.3 &  2.6 &  5 \\
\hline
\end{tabular}
\end{center}
}
\end{table*}

A first comparison of our individually derived metallicity estimates
shows that E02's assumption of a mean cluster metallicity of $0.4
Z_\odot$ (based on spectroscopic metallicity determinations by P93) is
reasonable, except for a few clusters with significantly supersolar
abundances.  Secondly, our low extinction values are also in line with
our expectations for the circumnuclear ring clusters; although the
extinction in the starburst ring varies in the range $1 \lesssim A_V
\lesssim 4$ mag (Grotheus \& Schmidt-Kaler 1991), is has been shown that
circumnuclear ring clusters are often either almost fully obscured or
virtually dust free (e.g., Maoz et al.  2001).  In addition, Meurer et
al.  (1995) estimate the average extinction in the NGC 3310 starburst
regions to be in the range $0.18 \lesssim {\rm E}(B-V) \lesssim 0.23$,
based on the slope of the UV SED and published Balmer decrement
measurements (P93), respectively, and P93 estimated $0.13 \lesssim {\rm
E}(B-V) \lesssim 0.37$ based on spectroscopic measurements (see also
S96, D00).  Since these extinction estimates cover both the YSCs and the
ISM, they are consistent with being upper limits to the extinction
towards individual YSCs. 

For their large-scale complexes, E02 assumed a mean age of $\sim 10^7$ yr,
which is somewhat underestimated compared to our mean age of the
individual clusters in these complexes of $\sim 4 \times 10^7$ yr.  For
the ``Jumbo'' region (E02 complex 108+109), P93 derived an age of $1.45
\times 10^7$ yr from spectrophotometry, which is consistent with the
median age we derive for the clusters in this region, $t_{\rm Jumbo}
\sim 1.3 \times 10^7$ yr.

Although the match between our individual cluster age estimates and the
E02 results is good for clusters 8 (E02 estimate $\sim 20-30$ Myr; our
estimate $\sim 26$ Myr), 19 ($\sim 20-30$ vs.  $\sim 21$ Myr) and 64
($10-50$ vs.  33 Myr), and reasonable for clusters 36 ($\lesssim 5-10$
vs.  $\sim 9-10$ Myr) and 90 ($10-50$ vs.  14 Myr), it is poor for
clusters 37 ($\sim 1-5$ vs.  $\sim 16$ Myr; this is no surprise, because
our models do not extend to such young ages), 49 ($10-50$ vs.  94 Myr),
and 59 ($10-50$ vs.  320 Myr).  Owing to the strong age-dependence of the M/L
ratio at those young ages, small differences in our age estimates {\it will} have
significant consequences for the final mass determinations, for ages in
the range determined for the NGC 3310 clusters.  In addition, the
difference in the assumed distance modulus to NGC 3310 between ourselves
($D = 13$ Mpc; Sect.  \ref{sources.sect}) and E02 ($D = 18.7$ Mpc) of
$\sim -0.8$ mag introduces an additional uncertainty to the clusters'
luminosity of $\Delta \log L \simeq 0.3$, corresponding to a factor of
$\sim 2$ in luminosity, and therefore in the mass (via the M/L ratio) as
well, in the sense that we expect the E02 mass estimates to be $\sim 2
\times$ higher. 

Thus, from a comparison between our final masses and those of E02, we
conclude that the factor of $2-3$ between most of our mass estimates, in
the sense that E02's masses are the higher, is not too surprising in
view of the many uncertainties involved in (i) the difference in
distance to NGC 3310 assumed, (ii) the different methods used for our
mass estimates, (iii) the different ways in which we obtained our
aperture photometry and the subsequent photometric calibrations (E02
converted their measurements to the standard ground-based system, while
we continued in the STMAG system), and (iv) the different models
employed (cf.  Fritze--v.  Alvensleben 2000).  Regarding the latter, we
point out that these differences may have been augmented by our
inclusion of nebular emission (which we have shown to be important for
ages younger than $\sim 3 \times 10^7$ yr; Anders et al.  2002, Anders
\& Fritze--v.  Alvensleben 2003) and our approach to solve for all of
our free parameters simultaneously. 

This comparison clearly shows that while we can obtain robust relative
distributions of masses, ages, and metallicities for a given star
cluster system fairly easily (as shown throughout this paper), the
absolute calibration, and therefore the absolute mass, age and
metallicity scales can only be fixed robustly if we have access to
independent {\it spectroscopic} measurements of these parameters. 

\subsection{The NGC 3310 Cluster Luminosity Function}

The importance of correcting the observed CLFs to a common age cannot be
overemphasized.  Age spread effects in young cluster systems, and
therefore the combined effects of ongoing cluster formation,
evolutionary fading, and the onset of cluster disruption, affect the
observed CLF (cf.  Meurer 1995, Fritze--v.  Alvensleben 1999, de Grijs
et al.  2001, 2003a,b).  This implies that the CLF observed in such a
system represents merely a temporal situation, rather than a
characteristic property of a coeval cluster system. 

Using the age estimates obtained for the individual clusters, we
corrected the F606W magnitudes of our cluster sample to a common age of
$10^{7.5}$ yr using the Anders \& Fritze--v.  Alvensleben (2003) models
properly folded through the {\sl HST/WFPC2} F606W filter response curve,
for the ``best-fit'' age distribution of Fig.  \ref{cffields.fig}a.  In
order to provide a comparison with the CLFs of other young cluster
systems, we employed the Levenberg-Marquardt method (Marquardt 1963) to
fit power-law slopes of the form $N(L) \propto L^{\alpha}$ (where $L$ is
the luminosity of a cluster, and $\alpha$ the power-law slope) to both
CLFs. 

In the {\it V}-band luminosity range $10^6 \le L_{\rm F606W}/L_\odot \le
10^7$ ($17.7 \lesssim {\rm F606W} \lesssim 20.2$ mag), the slope of the
CLF is $\alpha = -1.4 \pm 0.2$, and $\alpha = -1.8 \pm 0.4$ for the
full, uncorrected sample of clusters, and for the ``best-fit'' sample
corrected to a common age of $\log( {\rm age/yr} ) = 7.5$, respectively. 

In de Grijs et al.  (2001) and Parmentier et al.  (2002) we showed that
the constant-age {\it V}-band CLF for the bright clusters in M82's
post-starburst region ``B'' roughly follows a power law with a slope in
the range $-1.4 \lesssim \alpha \lesssim -1.2$, which is consistent with
the power-law slopes of other young CLFs, and within the uncertainties
also with the NGC 3310 CLF.  On the other hand, D00 found a
significantly shallower slope of $\alpha \sim -1$ for the
H$\alpha$-bright circumnuclear star forming regions in NGC 3310 (but we
note that these are likely the youngest objects populating the CLF, with
ages $\lesssim 6-12$ Myr, depending on metallicity), while E02 obtained
slopes as steep as $(-2.4 \pm 0.04) \le \alpha \le (-2.2 \pm 0.03)$ from
their {\it B} and {\it K}-band cluster measurements, which is -- within
the uncertainties -- consistent with our {\it V}-band determination. 
The main uncertainties that are expected to affect the CLF slope in this
latter case are the luminosity range over which the slopes were
obtained, and the fact that E02 did not correct their CLFs to a common
age before measuring their slopes. 

\section{Summary and Conclusions}
\label{summary.sect}

\subsection{Systematic uncertainties}

In this paper we have investigated the systematic uncertainties involved
in using broad-band UV, optical and NIR observations to derive age,
mass, metallicity and extinction distributions for extragalactic star
cluster systems.  The large majority of extragalactic star cluster
studies done to date have essentially used two or three-passband
aperture photometry, combined with theoretical stellar population
synthesis models to obtain age estimates.  The accuracy to which this
can be done obviously depends on the number of different (broad-band)
filters available as well as, crucially, on the actual wavelengths and
the wavelength range covered by the observations.  Understanding these
systematic uncertainties is therefore of the utmost importance for a
well-balanced interpretation of the properties of such star cluster
systems. 

We focused our analysis on the nearby, well-studied starburst galaxy NGC
3310, known to harbour large numbers of young star clusters, for which
we obtained multi-passband observations from the UV to the NIR from
the {\sl HST} Data Archive.  We used our evolutionary synthesis
optimisation technique to simultaneously determine the best combination
of age, extinction and metallicity. 

Our main results can be summarised as follows. For a star cluster system
as young as that in NGC 3310,

\begin{itemize}

\item the peak of the age distribution is robustly reproduced for all of
our choices of passband combinations, except for the red-only selected
filter combination;

\item red-dominated passband combinations result in significantly
different (uncertain) age solutions, due to the weak time dependence of
the NIR magnitudes and due to unavoidable ambiguities in the modelling
of the thermally pulsing AGB phase; they produce a significant wing of
older clusters. 

\item blue-selected passband combinations tend to result in age
estimates that are slightly skewed towards younger ages, compared to
passband combinations that also include redder passbands.  This is due
to the combination of observational selection effects and to the
age--metallicity degeneracy, which is also of importance in passband
combinations containing only optical filters. 

\item for ages $6 \lesssim \log( {\rm age/yr} ) \lesssim 9$, we conclude
that if one can only obtain partial coverage of a star cluster's SED, an
optical passband combination of at least 4 filters including {\it both}
blue {\it and} red optical passbands results in the best balanced and
most representative age distribution.  While blue-selected passband
combinations lead to age distributions that are slightly biased towards
younger ages, red and in particular NIR-dominated passband combinations
should be avoided. 

\item The physical effect most limiting the accuracy of our age
determinations, for ages up to $\sim 10^8$ yr, is the age--metallicity
degeneracy.  This is clearly illustrated in our comparison with results
obtained assuming a fixed, solar metallicity.  If the actual cluster
metallicities are significantly subsolar, this assumption results in a
markedly different age distribution, biased towards younger ages,
compared to the distribution obtained by leaving the metallicity as a
free parameter. 

\end{itemize}

\subsection{The NGC 3310 cluster system}

NGC 3310 is a local, very active starburst galaxy with high global star
formation efficiency, which has experienced very recent star formation
in star clusters and H{\sc ii} regions, in the last $\lesssim 10^7-10^8$
yr.  In the inner FoV covered by all of our passbands, we detected some
300 star cluster candidates, while an additional $\sim 100$ clusters
were detected in the outer FoV covered by the {\sl WFPC2} filters only,
all of them affected by only moderate levels of NGC 3310-internal
extinction towards the clusters, E$(B-V) \lesssim 0.1$ mag.  The age
distribution derived for these clusters indicates that NGC 3310
underwent a significant burst of cluster formation some $3 \times 10^7$
yr ago.  The actual duration of the burst of cluster formation, which we
estimate to have lasted for almost 20 Myr, may have been shorter because
uncertainties in the age determinations may have broadened the peak.  It
appears, therefore, that the peak of cluster formation in NGC 3310
coincides closely with the suspected galactic cannibalism or last tidal
interaction. 

The clusters older than $\sim 10^7$ yr are smoothly distributed
throughout the galactic centre.  However, the younger clusters, with
ages $\log( {\rm age/yr}) \le 7.1$, are predominantly concentrated in
the Jumbo region and in the northern spiral arm.  This is consistent
with the very young ages independently derived for these regions, $t
\lesssim 10$ Myr, although both regions also contain clusters spanning
the entire age range observed for the NGC 3310 star cluster system.  

The CLF is the result of a complex interplay of the intrinsic cluster
mass distribution, age spread and cluster disruption processes.  We
therefore determined the individual cluster masses by scaling our model
SEDs for the best-fit age, extinction values and metallicity estimates
to the observed SEDs.  We estimate the cluster system to have a median
mass of $\langle \log( m/M_\odot ) \rangle \sim 5.25 \pm 0.1$, not
including systematic uncertainties introduced by the uncertainties in
the low-mass IMF slope. 

Our metallicity determinations are strongly dominated by (significantly)
subsolar metallicities, which is consistent with independent metallicity
measurements.  There is some evidence that the most actively star
forming regions, in particular the Jumbo region and the northern spiral
arm, are predominantly composed of lower-abundance star clusters.  The
{\it V}-band CLF slope in the range $10^6 \le L_{\rm F606W}/L_\odot \le
10^7$ for the NGC 3310 star cluster system is $\alpha_{\rm F606W} \sim
-1.8 \pm 0.4$, which is consistent with the power-law slopes of other
young CLFs.

Finally, we point out that our estimates for the ages, masses,
metallicities and extinction values of the YSCs in NGC 3310 closely
match previous, independent determinations of these parameters, where
available (e.g., Grotheus \& Schmidt-Kaler 1991, P93, Meurer et al. 
1995, D00, E02).  This implies, to first order, that this type of
analysis is reasonably robust. 

\section*{Acknowledgments} This paper is based on new and archival
observations with the NASA/ESA {\sl Hubble Space Telescope}, obtained at
the Space Telescope Science Institute (STScI), which is operated by the
Association of Universities for Research in Astronomy (AURA), Inc.,
under NASA contract NAS 5-26555.  This paper is also partially based on
ASTROVIRTEL research support, a project funded by the European
Commission under 5FP Contract HPRI-CT-1999-00081.  This research has
made use of NASA's Astrophysics Data System Abstract Service.  PA is
partially supported by DFG grant Fr 916/11-1; PA also acknowledges
partial funding from the Marie Curie Fellowship programme EARASTARGAL
``The Evolution of Stars and Galaxies'', funded by the European
Commission under 5FP contract HPMT-CT-2000-00132.  VAT and RAW
acknowledge support from NASA grants GO-8645.*, awarded by STScI.  RdG
acknowledges preliminary analysis of the Cycle 9 {\sl HST} UV data by
Xsitaaz Chadee as part of the 2001 PPARC/Cambridge International
Undergraduate Summer School.

\end{document}